\documentclass{optica-article}

\journal{oe}
\articletype{Research Article}
\begin{document}

\title{Active Aerosols}

\author{Charles A. Rohde,\authormark{} Kristin M. Charipar,\authormark{} Paul Johns\authormark{}, Ashlin G. Porter,\authormark{} Nicholas J. Greybush, and Jake Fontana\authormark{*}}

\address{\authormark{}U.S. Naval Research Laboratory, 4555 Overlook Ave. SW, Washington, DC 20375, USA}

\email{\authormark{*}jake.fontana@nrl.navy.mil} %% email address is required; see note below about the corresponding author designation

% \homepage{http:...} %% author's URL, if desired

%%%%%%%%%%%%%%%%%%% abstract %%%%%%%%%%%%%%%%
%% [use \begin{abstract*}...\end{abstract*} if exempt from copyright]

\begin{abstract}
We report the dynamics and control of the orientational and positional order of ensembles of gold nanorods suspended in air at standard temperature and pressure using externally applied electric fields, demonstrating an active aerosol. Light filter, valve and gradient responses are shown, establishing active aerosols as a unique type of optical element we term component-less optics.
\end{abstract}

%%%%%%%%%%%%%%%%%%%%%%%%%%  body  %%%%%%%%%%%%%%%%%%%%%%%%%%
\section{Introduction}
The interaction between aerosols and light is a fundamental problem throughout optics. The consequences are wide ranging from influencing astronomical observations through cosmic dust \cite{Bohren1983,Schlegel}, to affecting the climate by altering earth’s albedo \cite{ALBRECHT,hansen}, and impairing vision in navigation \cite{Choi,Popoff}.  Although aerosols have ubiquitous impact, often limiting the ability to make observations, they are commonly regarded as disordered media. Understanding the relationships that define order in aerosols is key to governing their optical properties.

Individual aerosol particles are routinely ordered using external fields \cite{Ashkin1970,Jauffred, Reimann,Conangla,smalley,shvedov}.  However, control of the macroscale optics of an aerosol ensemble has yet to be realized, generally due to the need for substantial coupling of an external field to a large number of particles (>billions) in low pressure environments to overcome the randomizing thermal forces of Brownian motion.

Plasmonic nanoparticles couple strongly to electromagnetic fields, providing a mechanism, material and opportunity to guide the response of an ensemble \cite{Prodan, miller2016fundamental, zhang2014, yang2016, avitzour2009, miller2022}.  Active plasmonics have further enabled the spatial, spectral and temporal ordering of these material properties with external stimuli \cite{Jiang, Shaltouteaat,salandrino2018plasmonic}. Recently, suspensions of gold nanorods were  transitioned from the liquid to gas state and the isotropic optical response was characterized \emph{in~situ}, demonstrating a thermodynamically stable and optically homogeneous plasmonic aerosol \cite{geldmeier}.  Yet, to date, the dynamic anisotropic response of these materials in air when placed in external fields remains unknown.

\begin{figure}[htp]
\centering\includegraphics[width=7cm]{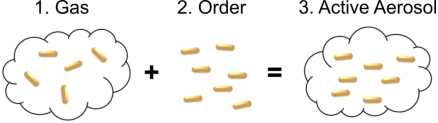}
\vspace{0cm}
\caption{An active aerosol modulates light by controlling the order of particles in a gas using external fields.  }
\label{fig:setup}
\end{figure}

In this paper,  we propose and demonstrate an active aerosol by controlling ensembles of gold nanorods suspended in air by varying magnitude and frequency of an externally applied electric field (Fig. 1).  We utilize active aerosols to establish the concept of component-less optics using three illustrative paradigms to traditional optics: (i) The isotropic response of the aerosol behaves as a stop-band filter, absorbing discrete wavelength bands. (ii) A light valve response is shown by directing the orientational order of the nanorods in air  with electric fields, temporally opening or closing spectral bands and polarization states.  (iii) The spatial distribution of the aerosolized nanorods was varied by electrophoretic force, providing a mechanism to tailor the gradient-index profile within the aerosol.

\section{Results and discussion}

%\subsection{Experimental setup}

Experiments were carried out to understand the dynamics of gold nanorods suspended in air when external electric fields are applied. Gold nanorods with a length to diameter aspect ratio of 5.7~(114~nm/20~nm) were synthesized to overlap the longitudinal absorption peak of the nanorods in air with the probe laser wavelength (1030~nm) \cite{Park}. Aqueous suspensions of nanorods initially stablized with CTAB (hexadecyltrimethylammonium bromide) were phase transferred into ethyl acetate by coating the surface of the nanorods with thiol terminated polystyrene (Polymer Source, Inc., $M_{n}=29k$, $M_{w}=31k$) \cite{fontana}, allowing for rapid evaporation of the solvent droplets upon aerosolization. The liquid suspension of nanorods (40~mL, 0.5~nM) were aerosolized into the gas phase, creating a plasmonic aerosol, using a Venturi based atomizer \cite{geldmeier}.

The plasmonic aerosol at standard temperature and pressure (nanorod volume fraction, $\phi=4.90\times10^{-10}$) filled a probe cell (3D printed polyamide, cross-section = 1000~mm$^2$, length = 150~mm) from the bottom port (Fig. \ref{fig:setup}).  A pair of rectangular copper electrodes (length = 150~mm, height = 12~mm, spacing = 3~mm) along the long axis of the probe cell applied a uniform electric field across the aerosol. 

\begin{figure}[htp]
\centering\includegraphics[width=9cm]{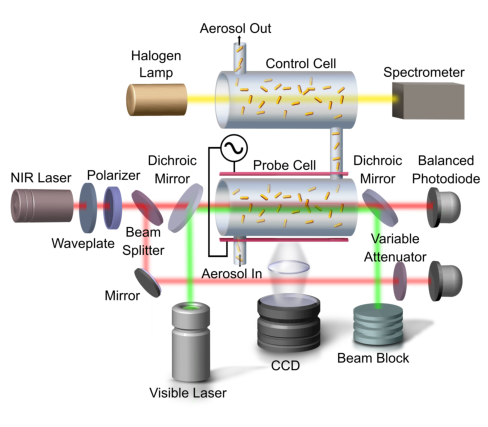}
\vspace{0cm}
\caption{Schematic of the experimental apparatus used to measure the evolution of the plasmonic aerosols.}
\label{fig:setup}
\end{figure}

\begin{figure}[htp]
\centering\includegraphics[width=8.5cm]{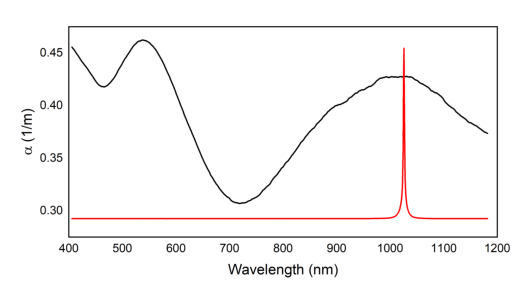}
\vspace{0cm}
\caption{Active aerosol light filter. A representative spectrum of a stop-band filter, consisting of an isotropic plasmonic aerosol and NIR probe laser.}
\label{fig:lightvalve}
\end{figure}

Upon exiting the probe cell from the top port, the aerosol flowed into a control cell, with identical dimensions to the probe cell. The aerosol spectrum was collected (tungsten halogen unpolarized light, Ocean Optics HL-2000-HP, QE-Pro and NIRQuest) and used to determine the absolute magnitude of the absorption (Fig. \ref{fig:lightvalve}). The aerosol passed through a filter (HEPA) removing the nanorods from the air after exiting the control cell.

\subsection{Light filter}

The isotropic optical response of the aerosol is shown in Figure 3.  The longitudinal and transverse absorption peaks of the nanorods in air were spectrally resolved at 1020 nm and 520 nm, respectively, showing a stop-band light filter response at the absorption peak wavelengths.  The stop- or pass-bands can be further tailored by tuning the aspect ratio of the nanorods \cite{geldmeier}.  The large transverse absorption peak, relative to liquid suspensions, is due to increased Rayleigh scattering from the polymer coated nanorods in air \cite{Cox}.

\subsection{Light valve}

To demonstrate control of the orientational order of an ensemble of gold nanorods suspended in air and resulting active aerosol light valve, a collimated (RC02FC-P01) near infrared laser diode source (QFLD-1030-10S, beam diameter = 2~mm, 0.5~mW diode laser) was incident upon a $1/2$-waveplate and polarizer, creating a linearly polarized beam parallel to the applied electric field in the probe cell. Using a 50:50 beamsplitter, half of the light irradiated the aerosol in the probe cell and the other half was used as a reference beam. The light exiting the probe cell was collected on the negative sensor of the balanced photodiode (BPD210A) and the reference beam was collected on the positive sensor.  The balanced photodiode voltage difference was adjusted to zero with a variable neutral density attenuator prior to the aerosols being introduced into the probe cell. 

\begin{figure}[htp]
\centering\includegraphics[width=13.5cm]{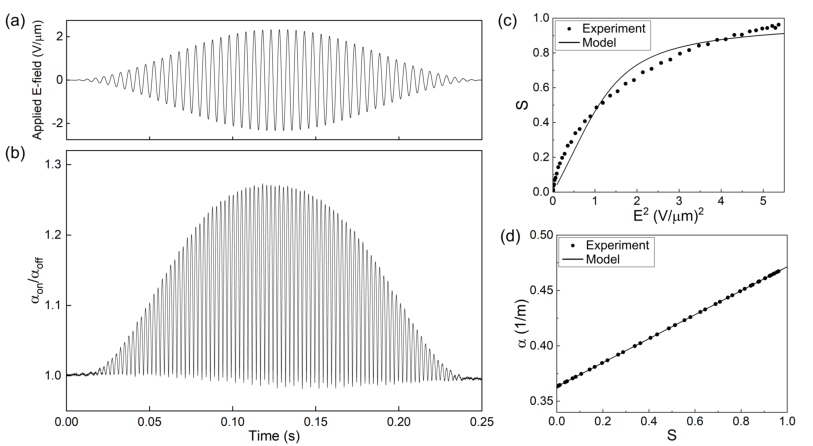}
\vspace{0cm}
\caption{Active aerosol light valve.  (a) The applied sine-envelope electric field pulse (50 cycle, 200~Hz, 0 to 2.3~$V/\mu m$). (b) The dynamic response of the relative absorption spectrum for gold nanorods suspended in air as a function of applied electric field. (c) Order parameter versus the square of the applied electric field.  The solid black lines are the calculated fits from the experimentally determined $A=0.108~m^{-1}$, $B=0.363~m^{-1}$, and $E_{c}=0.560~V/\mu m$ parameters. (d) Absorption versus order parameter.}
\label{fig:resultsA}
\end{figure}

The intensity difference between the balanced photodiode signals directly determined the relative change in the absorption, $\alpha_\mathrm{on}/\alpha_\mathrm{off}$, of the aerosols as a function of an applied electric field pulse (Fig. \ref{fig:resultsA}(a)). At 1030~nm the out of beam scattering is small, therefore the extinction is assumed to be dominated by absorption. The measurements of absorption at 1030~nm (Fig. \ref{fig:lightvalve}) were then used to calibrate $\alpha_\mathrm{on}$/$\alpha_\mathrm{off}$ from relative to absolute changes in magnitude.

For the experimental geometry, as the 1020~nm longitudinal absorption axis of the nanorods begin to align in the applied electric field and probe laser polarization direction, an increase in the relative absorption at 1030~nm is measured (Fig. \ref{fig:resultsA}(b)).

The absorption of the aerosol depends linearly on the orientational order of the nanorods \cite{fontana},
\begin{equation}
\alpha=AS+B,
\end{equation}
where $A=\frac{4\pi \phi}{3\lambda n_{h}}\left(Im[\chi_{\parallel}]-Im[\chi_{\perp}]  \right)$, $B=\frac{2\pi \phi}{3\lambda  n_{h}}\left(Im[\chi_{\parallel}]+2Im[\chi_{\perp}]  \right)$, $\lambda$ is the probe wavelength, $n_{h}$ is the refractive index of the host, $Im[\chi_{\parallel, \perp}]$ are the imaginary parts of the longitudinal and transverse principal values of the nanorod susceptibility and $S$ is the scalar orientational order parameter,

\begin{equation}
S=\frac{\int_{0}^{1}\frac{1}{2}\left(3cos^2\theta-1\right)e^{\frac{E^2}{E^2_{c}}cos^2\theta} \,d(cos\theta)}{\int_{0}^{1}e^{\frac{E^2}{E^2_{c}}cos^2\theta} \,d(cos\theta)},
\end{equation}

where $\theta$ is angle between the applied electric field direction and the longitudinal axis of the nanorod.  The critical magnitude of the electric field needed for alignment is $E_{c}=\sqrt{2k_{B}T/\left(\left(\frac{1}{L_{\parallel}}-\frac{1}{L_{\perp}}\right)\epsilon_{0}\epsilon_{h}V\right)}$, where $k_{B}$ is Boltzmann's constant, $T$ is absolute temperature, $L_{\parallel,\perp}$ are the depolarization factors for fields parallel and perpendicular to the long axis of the nanorod, $\epsilon_{0,h}$ are the permittivites of free space and host material and $V$ is the volume of a nanorod.

The measurement of $\alpha$ versus E was fit to Eq. (1), with fitting parameters A, B and $E_{c}$ (Fig. \ref{fig:resultsA}(c,d)).   The solid black lines are the calculated fits from the fitted parameters. The minor evolution difference between the model and experimental data in Fig. 4(c) is likely due to a non-negligible frequency-dependent translation of the nanorods, as evidenced by the slight deviation from unity for large electric field magnitudes in Fig. 4(b).  We find the absorption of the aerosol is linear in the order parameter (Fig. 4(d)), verifying the relationship in Eq. (1). The order parameter maximum is $S_{max}=0.96$.  The magnitude of $E_{c}=0.560~V/\mu m$ and ensuing potential energy of the nanorod, relative to the thermal energy, is expected for nanorod alignment \cite{fontana}. 

Inverting the $A$ and $B$ parameters and solving for the imaginary principal values of the nanorod susceptibility at 200~Hz yields: $Im[\chi_{\parallel}]=164$, $Im[\chi_{\perp}]=107$. The low-frequency susceptibility is anticipated to be several orders of magnitude larger than the optical susceptibility, assuming the imaginary permittivity, $Im[\epsilon]=\sigma /\omega$, increases as $\omega\rightarrow0$, \cite{Olmon} if the conductivity of gold, $\sigma$, is assumed constant.

\subsection{Gradient-index}

To guide the positional order of an ensemble of gold nanorods suspended in air, and to show an active aerosol gradient response, a DC electric field was applied across the electrodes in the probe cell (Fig. 5).  The scattered light from the nanorods was imaged with a 4x microscope objective (Nikon MRP00042) onto a CMOS camera (Thorlabs DCC3240N, 10 Hz) through a window inserted in the top of the probe cell.  A second diode laser (QFLD-520-10S) at 520~nm wavelength was co-propagated, using dichroic mirrors, along the NIR beam axis through the probe cell to increase the sensitivity of the CCD measurement (Fig. 2).

\begin{figure}[htbp]
%\label{Figure3:DCres}
\centering\includegraphics[width=13.5cm]{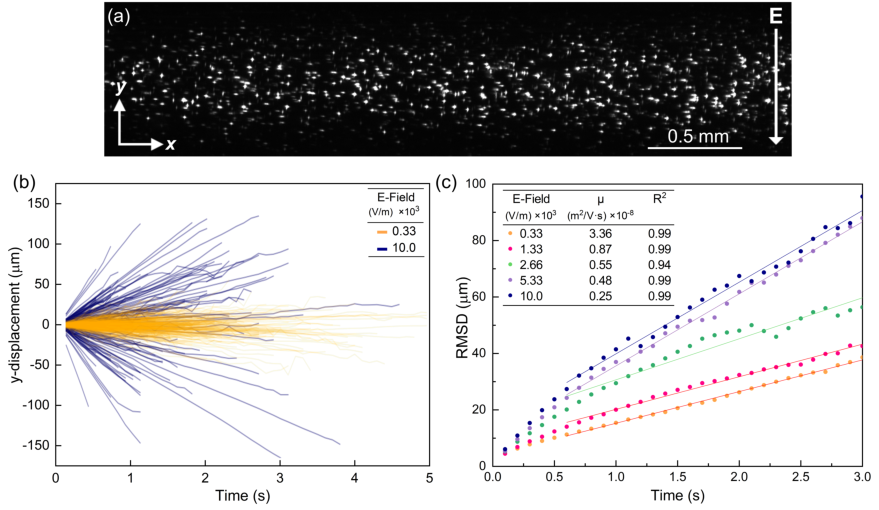}
\vspace{-.5cm}
\caption{Active aerosol gradient.  (a) Representative image of the plasmonic aerosol. Evolution of the vertical displacement (b) and ensemble root mean square displacement (c)  of the aerosol versus the magnitude of the DC electric field. The inset table is the fitted values of the electrophoretic mobility.}

\end{figure}

In Fig. 5(a) we present an image of the scattered light from the nanorods suspended in air.  A DC pulse ($3~s$ on, $30~s$ off) was applied across the aerosol.  To ensure the images continuously captured the translation of the nanorods in the relatively small volume between the electrodes, small voltages ($0-30~V$) were applied to minimize the electrophoretic force and resulting nanorod velocities.  Software was used to analyze the motion of the nanorods between image frames \cite{trakpy}.

As the magnitude of the DC electric field increased, the average displacement velocity of the nanorods increased in the applied field direction ($y$-axis) (Fig. 5(b)).  The nanorods translated in both the plus and minus field directions due to residual surfactant encased in the ligand shell coating the nanorods, leading to a net positive or negative charge on the nanorods and the bipolar translation along the field direction. The nanorod translation along the aerosol flow direction ($x$-axis) was constant and unaffected by the electric field.

The ensemble root mean square displacement (RMSD) of the nanorod aerosol depends linearly on the magnitude of the applied electric field (Fig. 5(c)). The datasets are fitted using the ensemble RMSD,

\begin{equation}
\sqrt{ \frac{1}{N}\sum_{i=0}^{N} ( y_{i}(t)-y_{i}(0))^{2} }= \mu Et,
\end{equation}

where $N$ is the number of nanorods, $\mu$ is the electrophoretic mobility and $y_{i}(t)$ is the $y$-position of a nanorod at time, $t$.  The model was fit at different electric field magnitudes and in the constant velocity regime $t=0.6-3~s$.  An average ensemble electrophoretic mobility of $\bar\mu =1.1\pm 1.3\times10^{-8}~m^{2}/(V\cdot s)$ was retrieved from the fits.  From the mobility, the magnitude of the charge on the nanorods is $\lvert \bar q\rvert =6\pi \eta r \bar\mu=0.83\pm1.2\times10^{-19}~C$, where $\eta$ is the viscosity of air and $r$ is the hydrodynamic radius of the nanorod.  The resulting value of $\bar q$ is within experimental reason, if each nanorod acquires a net plus or minus charge upon aerosolization.

The positional order of the ensemble is directed by the electric field induced density variation in the aerosol, resulting in a refractive index gradient, $\partial n/\partial y \approx 10^{-6}$, where $n \approx 1+\frac{1}{2}\phi \chi$.  While the experimental magnitude of $\phi \approx 10^{-10}$ leads to an exceedingly small refractive index gradient, the mechanism to direct the positional order of the aerosol is apparent. 

Further investigations may optimize the orientational and positional ordering mechanisms described above and possibly lead to additional component-less optics such as diffusers, lenses or holographic elements, as shown in Fig. 6 \cite{palmer1980, palmer1983}.  If multiple component-less optics are combined, then integrated optical circuits may be realized, opening the possibility for the generation, propagation, detection and processing of light using active aerosols \cite{Engheta, Silva}. 

\begin{figure}[htbp]
%\label{Figure3:DCres}
\centering\includegraphics[width=9cm]{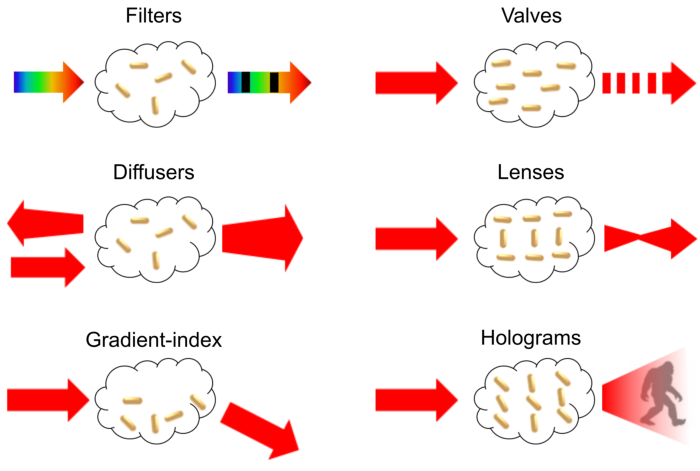}
\vspace{0cm}
\caption{Illustrative examples of component-less optic elements.}
\label{fig:opticalelements}
\end{figure}

\section{Conclusion}

We have experimentally demonstrated an active aerosol by controlling the order of gold nanorods suspended in air using electric fields.  We applied active aerosols to realize component-less optics and highlighted three illustrative paradigms.  These results open up the possibility to govern electrodynamic effects in the atmosphere.

\begin{backmatter}

\bmsection{Acknowledgments}
J.F. thanks Rohith Chandrasekar and Richard Vaia for insightful discussions. This work was supported by the Office of Naval Research (N0001421WX00025, N0001422WX00600).

\bmsection{Disclosures}
The authors declare no conflicts of interest.

\bmsection{Data availability} Data underlying the results presented in this paper are not publicly available at this time but may be obtained from the authors upon reasonable request.

\end{backmatter}

%%%%%%%%%%%%%%%%%%%%%%% References %%%%%%%%%%%%%%%%%%%%%%%%%

%%%%%%%%%% If using BibTeX:

\bibliography{references}

\begin{thebibliography}{10}
\newcommand{\enquote}[1]{``#1''}

\bibitem{Bohren1983}
C.~F. Bohren and D.~R. Huffman, \emph{Absorption and Scattering of Light by
  Small Particles} (Wiley-VCH, New York, NY, 1983).

\bibitem{Schlegel}
D.~J. Schlegel, D.~P. Finkbeiner, and M.~Davis, \enquote{Maps of dust infrared
  emission for use in estimation of reddening and cosmic microwave background
  radiation foregrounds,} {\protect\JournalTitle{The Astrophysical Journal}}
  \textbf{500}, 525--553 (1998).

\bibitem{ALBRECHT}
B.~A. Albrecht, \enquote{Aerosols, cloud microphysics, and fractional
  cloudiness,} {\protect\JournalTitle{Science}} \textbf{245}, 1227--1230
  (1989).

\bibitem{hansen}
J.~Hansen and L.~Nazarenko, \enquote{Soot climate forcing via snow and ice
  albedos,} {\protect\JournalTitle{Proc. Natl. Acad. Sci. U.S.A.}}
  \textbf{101}, 423--428 (2004).

\bibitem{Choi}
Y.~Choi, T.~D. Yang, C.~Fang-Yen, P.~Kang, K.~J. Lee, R.~R. Dasari, M.~S. Feld,
  and W.~Choi, \enquote{Overcoming the diffraction limit using multiple light
  scattering in a highly disordered medium,} {\protect\JournalTitle{Phys. Rev.
  Lett.}} \textbf{107}, 023902 (2011).

\bibitem{Popoff}
S.~M. Popoff, G.~Lerosey, R.~Carminati, M.~Fink, A.~C. Boccara, and S.~Gigan,
  \enquote{Measuring the transmission matrix in optics: An approach to the
  study and control of light propagation in disordered media,}
  {\protect\JournalTitle{Phys. Rev. Lett.}} \textbf{104}, 100601 (2010).

\bibitem{Ashkin1970}
A.~Ashkin, \enquote{Acceleration and trapping of particles by radiation
  pressure,} {\protect\JournalTitle{Phys. Rev. Lett.}} \textbf{24}, 156--159
  (1970).

\bibitem{Jauffred}
L.~Jauffred, S.~M.-R. Taheri, R.~Schmitt, H.~Linke, and L.~B. Oddershede,
  \enquote{Optical trapping of gold nanoparticles in air,}
  {\protect\JournalTitle{Nano Letters}} \textbf{15}, 4713--4719 (2015).

\bibitem{Reimann}
R.~Reimann, M.~Doderer, E.~Hebestreit, R.~Diehl, M.~Frimmer, D.~Windey,
  F.~Tebbenjohanns, and L.~Novotny, \enquote{Ghz rotation of an optically
  trapped nanoparticle in vacuum,} {\protect\JournalTitle{Phys. Rev. Lett.}}
  \textbf{121}, 033602 (2018).

\bibitem{Conangla}
G.~P. Conangla, R.~A. Rica, and R.~Quidant, \enquote{Extending vacuum trapping
  to absorbing objects with hybrid paul-optical traps,}
  {\protect\JournalTitle{Nano Letters}} \textbf{20}, 6018--6023 (2020).

\bibitem{smalley}
D.~Smalley, E.~Nygaard, K.~Squire, J.~Van~Wagoner, J.~Rasmussen, S.~Gneiting,
  K.~Qaderi, J.~Goodsell, W.~Rogers, M.~Lindsey, K.~Costner, A.~Monk,
  M.~Pearson, B.~Haymore, and J.~Peatross, \enquote{A photophoretic-trap
  volumetric display,} {\protect\JournalTitle{Nature}} \textbf{553}, 486--490
  (2018).

\bibitem{shvedov}
V.~G. Shvedov, A.~V. Rode, Y.~V. Izdebskaya, A.~S. Desyatnikov, W.~Krolikowski,
  and Y.~S. Kivshar, \enquote{Giant optical manipulation,}
  {\protect\JournalTitle{Phys. Rev. Lett.}} \textbf{105}, 118103 (2010).

\bibitem{Prodan}
E.~Prodan, C.~Radloff, N.~J. Halas, and P.~Nordlander, \enquote{A hybridization
  model for the plasmon response of complex nanostructures,}
  {\protect\JournalTitle{Science}} \textbf{302}, 419--422 (2003).

\bibitem{miller2016fundamental}
O.~D. Miller, A.~G. Polimeridis, M.~H. Reid, C.~W. Hsu, B.~G. DeLacy, J.~D.
  Joannopoulos, M.~Solja{\v{c}}i{\'c}, and S.~G. Johnson, \enquote{Fundamental
  limits to optical response in absorptive systems,}
  {\protect\JournalTitle{Opt. Express}} \textbf{24}, 3329--3364 (2016).

\bibitem{zhang2014}
H.~Zhang, H.~V. Demir, and A.~O. Govorov, \enquote{Plasmonic metamaterials and
  nanocomposites with the narrow transparency window effect in broad extinction
  spectra,} {\protect\JournalTitle{ACS Photonics}} \textbf{1}, 822--832 (2014).

\bibitem{yang2016}
J.~Yang, N.~J. Kramer, K.~S. Schramke, L.~M. Wheeler, L.~V. Besteiro, C.~J.
  Hogan~Jr, A.~O. Govorov, and U.~R. Kortshagen, \enquote{Broadband absorbing
  exciton--plasmon metafluids with narrow transparency windows,}
  {\protect\JournalTitle{Nano letters}} \textbf{16}, 1472--1477 (2016).

\bibitem{avitzour2009}
Y.~Avitzour, Y.~A. Urzhumov, and G.~Shvets, \enquote{Wide-angle infrared
  absorber based on a negative-index plasmonic metamaterial,}
  {\protect\JournalTitle{Physical Review B}} \textbf{79}, 045131 (2009).

\bibitem{miller2022}
O.~Miller, K.~Park, and R.~A. Vaia, \enquote{Towards maximum optical efficiency
  of ensembles of colloidal nanorods,} {\protect\JournalTitle{Optics Express}}
  \textbf{30}, 25061--25077 (2022).

\bibitem{Jiang}
N.~Jiang, X.~Zhuo, and J.~Wang, \enquote{Active plasmonics: Principles,
  structures, and applications,} {\protect\JournalTitle{Chemical Reviews}}
  \textbf{118}, 3054--3099 (2018).

\bibitem{Shaltouteaat}
A.~M. Shaltout, V.~M. Shalaev, and M.~L. Brongersma, \enquote{Spatiotemporal
  light control with active metasurfaces,} {\protect\JournalTitle{Science}}
  \textbf{364}, eaat3100 (2019).

\bibitem{salandrino2018plasmonic}
A.~Salandrino, \enquote{Plasmonic parametric resonance,}
  {\protect\JournalTitle{Physical Review B}} \textbf{97}, 081401 (2018).

\bibitem{geldmeier}
J.~Geldmeier, P.~Johns, N.~J. Greybush, J.~Naciri, and J.~Fontana,
  \enquote{Plasmonic aerosols,} {\protect\JournalTitle{Physical Review B}}
  \textbf{99}, 081112 (2019).

\bibitem{Park}
K.~Park, M.-s. Hsiao, Y.-J. Yi, S.~Izor, H.~Koerner, A.~Jawaid, and R.~A. Vaia,
  \enquote{Highly concentrated seed-mediated synthesis of monodispersed gold
  nanorods,} {\protect\JournalTitle{ACS Applied Materials \& Interfaces}}
  \textbf{9}, 26363--26371 (2017).

\bibitem{fontana}
J.~Fontana, G.~K. da~Costa, J.~M. Pereira, J.~Naciri, B.~R. Ratna,
  P.~Palffy-Muhoray, and I.~C. Carvalho, \enquote{Electric field induced
  orientational order of gold nanorods in dilute organic suspensions,}
  {\protect\JournalTitle{Applied Physics Letters}} \textbf{108}, 081904 (2016).

\bibitem{Cox}
A.~J. Cox, A.~J. DeWeerd, and J.~Linden, \enquote{An experiment to measure mie
  and rayleigh total scattering cross sections,}
  {\protect\JournalTitle{American Journal of Physics}} \textbf{70}, 620--625
  (2002).

\bibitem{Olmon}
R.~L. Olmon, B.~Slovick, T.~W. Johnson, D.~Shelton, S.-H. Oh, G.~D. Boreman,
  and M.~B. Raschke, \enquote{Optical dielectric function of gold,}
  {\protect\JournalTitle{Phys. Rev. B}} \textbf{86}, 235147 (2012).

\bibitem{trakpy}
D.~Allan, T.~A. Caswell, N.~Keim, F.~Boulogne, R.~W. Perry, and L.~Uieda,
  \enquote{Soft matter trackpy v0.5,}  (2021).

\bibitem{palmer1980}
A.~J. Palmer, \enquote{Nonlinear optics in aerosols,}
  {\protect\JournalTitle{Optics Letters}} \textbf{5}, 54--55 (1980).

\bibitem{palmer1983}
A.~Palmer, \enquote{Plasmon-resonant aerosols for space optics,}
  {\protect\JournalTitle{JOSA}} \textbf{73}, 1568--1573 (1983).

\bibitem{Engheta}
N.~Engheta, A.~Salandrino, and A.~Al\`u, \enquote{Circuit elements at optical
  frequencies: Nanoinductors, nanocapacitors, and nanoresistors,}
  {\protect\JournalTitle{Phys. Rev. Lett.}} \textbf{95}, 095504 (2005).

\bibitem{Silva}
A.~Silva, F.~Monticone, G.~Castaldi, V.~Galdi, A.~Al{\`u}, and N.~Engheta,
  \enquote{Performing mathematical operations with metamaterials,}
  {\protect\JournalTitle{Science}} \textbf{343}, 160--163 (2014).

\end{thebibliography}

%\end{references}

\end{document}